\documentclass[sigconf]{acmart}

\usepackage{subcaption}
\usepackage{enumitem}
\usepackage{algorithm}
\usepackage{graphicx}
\usepackage[noend]{algorithmic}
\usepackage{tabularx}
\usepackage{multirow}
\usepackage{makecell}
\usepackage[normalem]{ulem}

\usepackage{bm}

\AtBeginDocument{%
  \providecommand\BibTeX{{%
    \normalfont B\kern-0.5em{\scshape i\kern-0.25em b}\kern-0.8em\TeX}}}

\copyrightyear{2024}
\acmYear{2024}
\setcopyright{rightsretained}
\acmConference[CIKM '24]{Proceedings of the 33rd ACM International Conference on Information and Knowledge Management}{October 21--25, 2024}{Boise, ID, USA}
\acmBooktitle{Proceedings of the 33rd ACM International Conference on Information and Knowledge Management (CIKM '24), October 21--25, 2024, Boise, ID, USA}
\acmDOI{10.1145/3627673.3679662}
\acmISBN{979-8-4007-0436-9/24/10}

\makeatletter
\gdef\@copyrightpermission{
  \begin{minipage}{0.3\columnwidth}
   \href{https://creativecommons.org/licenses/by/4.0/}{\includegraphics[width=0.90\textwidth]{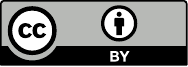}}
  \end{minipage}\hfill
  \begin{minipage}{0.7\columnwidth}
   \href{https://creativecommons.org/licenses/by/4.0/}{This work is licensed under a Creative Commons Attribution International 4.0 License.}
  \end{minipage}
  \vspace{5pt}
}
\makeatother

\begin{document}

\title{City Foundation Models for Learning General Purpose Representations from OpenStreetMap}


\author{Pasquale Balsebre}
\affiliation{
  \institution{Nanyang Technological University}
  \orcid{0009-0004-9454-2704}
  \country{Singapore}
}
\email{pasquale001@e.ntu.edu.sg}

\author{Weiming Huang}
\affiliation{
  \institution{Nanyang Technological University}
  \country{Singapore}}
\email{weiming.huang@nateko.lu.se}

\author{Gao Cong}
\affiliation{
  \institution{Nanyang Technological University}
  \country{Singapore}}
\email{gaocong@ntu.edu.sg}

\author{Yi Li}
\affiliation{
  \institution{Nanyang Technological University}
  \country{Singapore}}
\email{liyi0067@e.ntu.edu.sg}


\begin{abstract}
Pre-trained Foundation Models (PFMs) have ushered in a paradigm-shift in AI, due to their ability to learn general-purpose representations that can be readily employed in downstream tasks. While PFMs have been successfully adopted in various fields such as NLP and Computer Vision, their capacity in handling geospatial data remains limited. This can be attributed to the intrinsic heterogeneity of such data, which encompasses different types, including points, segments and regions, as well as multiple information modalities. The proliferation of Volunteered Geographic Information initiatives, like OpenStreetMap, unveils a promising opportunity to bridge this gap. In this paper, we present CityFM, a self-supervised framework to train a foundation model within a selected geographical area. CityFM relies solely on open data from OSM, and produces multimodal representations, incorporating spatial, visual, and textual information. We analyse the entity representations generated by our foundation models from a qualitative perspective, and conduct experiments on road, building, and region-level downstream tasks. In all the experiments, CityFM achieves performance superior to, or on par with, application-specific algorithms.


\end{abstract}





\maketitle

\section{Introduction}

The past decade has witnessed a shift from the exclusive provision of geospatial data from national mapping agencies, to freely available, volunteer-based sources. Researchers and practitioners from different disciplines have increasingly recognised the enormous potential of Volunteered Geographic Information (VGI) initiatives, and this trend has been observed in fields such as GIScience, urban planning, cartography and computer science, among others \cite{OSM_book}. OpenStreetMap (OSM), one of the most successful VGI projects, aims at building and maintaining a free, editable map of the world. 
Researchers have harnessed OSM data to train a plethora of algorithms for various tasks, such as traffic analysis \cite{rural_speed_prediction_osm, sigmod_travel_time}, land use prediction \cite{opensentinelmap} and recommendation \cite{next_poi_recommendation}. A typical approach, in the geospatial domain, is to design task-specific algorithms for each downstream application. This presents two major limitations: (1) the models require a large number of labeled samples for training, and (2) the models and representations learned for one task are not necessarily useful for other tasks. 

A promising solution to alleviate these limitations lies in adopting Pre-trained Foundation Models (PFMs). 
A key advantage of PFMs is that the pre-training phase is carried out self-supervisedly, without the need of human annotations; this allows the model to access larger amounts of data and produce effective representations that generalize across tasks. Given the scarcity of labeled geospatial data, and the diverse range of applications it serves, the adoption of PFMs in the geospatial domain presents promising opportunities \cite{geoai_survey_weiming}. However, such adoption is a non-trivial process, due to the fact that geospatial objects are characterized by multi-modal information, including a position in the space, textual annotations, and physical attributes such as shape and size. In addition, geospatial data exhibit inherent heterogeneity, requiring different approaches to handle the diverse entity types. For instance, OpenStreetMap's database stores spatial entities categorized into three types: \texttt{Nodes} to represent Points of Interest (POIs); \texttt{Ways} are multi-point geometries, comprising polylines and polygons, and are used to represent roads, bridges or large POIs; \texttt{Relations} are lists of \texttt{Nodes}, \texttt{Ways}, or \texttt{Relations}, called members, and represent relationships between them. Additionally, each entity can be optionally associated with a set of textual \texttt{Tags}, stored as key-value pairs.

Existing studies have primarily focused on a single data type, 
e.g., entities of type \texttt{Way} (polylines), to build a road network for traffic speed prediction \cite{IRN2Vec, RFN_speed_prediction}. While this approach is intuitive, it fails to leverage the diverse and multimodal nature of geospatial data types, which convey unique information aspects of the same entity. For example, a large portion of buildings, in OSM, is represented as polygons, but only $\sim$20\% is associated with a tag describing its functionality. Previous research works \cite{ambiguity_and_plausibility, osm_pois_urban_change, hex2vec} frequently omitted \textit{untagged} spatial entities, considered less informative; yet an important hint on an entity's functionality can be provided by auxiliary information, such as its shape, size and position. Other studies have used supplementary data of different modalities, including human trajectories \cite{toast} and street view images \cite{urban2vec}, to complement OSM data and enhance their framework's performance. However, such data may be expensive to obtain, or available only in specific cities. 

In this study, we propose \texttt{CityFM}, a framework to train a foundation model within a selected geographical area of interest, such as a city, capable of producing meaningful representations for geospatial entities of different types, which can be easily employed in a wide range of downstream tasks. In adherence to the best practices of data ownership and reproducibility, we design \texttt{CityFM} to rely solely on OSM data, which is freely available and accessible globally. \texttt{CityFM} is designed as a self-supervised learning architecture, based on mutual information maximization, capable of capturing the textual, visual and spatial characteristics of an entity. Computing effective and meaningful representations for OSM entities is especially challenging, due to the diversity of the available data and the different skills, tools and annotation styles of the contributors, which lead to sparsely and heterogeneously annotated entities. In fact, 
OSM promotes the so-called \textit{Any tags you like} 
policy, that lets annotators create their own keys and values for tags of entities they are adding. \texttt{CityFM} handles spatial objects of different types and is able to infer missing values, using other aspects of the same object, or entities in the spatial context. Furthermore, previous research works \cite{GeoVectors, toast, urban2vec, hex2vec, mining_geospatial_rel_from_text, osm_sigmod_1} have often overlooked \texttt{Relations}, owing to the very distinctive types of information represented. We introduce \texttt{Relations} in our self-supervised framework, showing they incorporate information about connectivity and public transports, and can guide the model to recognise important transportation hubs and arterial roads in a city. Finally, we conduct qualitative analyses, and quantitative experiments on a set of road, building and region-level downstream tasks, to demonstrate the utility of the representations produced by our foundation models. In conclusion:

\begin{itemize}[leftmargin=*]
  \item We propose a new framework, \texttt{CityFM}, for general-purpose representation learning in the geospatial domain, using exclusively volunteered multimodal data from OpenStreetMap.
  \item We design a self-supervised task, based on mutual information maximization, that introduces \texttt{Node}s, \texttt{Way}s and \texttt{Relation}s in the learning framework, to train a geospatial PFM within a selected area of interest. \texttt{CityFM}-trained models are multimodal and can integrate an entity's textual, visual and spatial aspects.
  \item We conduct experiments on road, building and region-level downstream tasks to showcase the effectiveness of the embeddings produced by our foundation models, compared to algorithms tailored for the specific tasks. The code, data, and pre-trained are made available\footnote{https://github.com/PasqualeTurin/CityFM}.
\end{itemize}

\section{Impact and Applications}
\texttt{CityFM} is a framework to pre-train a foundation model, without the need of human supervision, in a region of any shape and size. It stands out as the most comprehensive representation learning framework for geospatial vector data. 
CityFM’s comprehensiveness in encoding various geospatial entities shapes the model to be highly effective. 
We chose to utilize OpenStreetMap data for pre-training, due to the fact that it is the most complete, open source, geospatial database. 
Nonetheless, different data sources, with similar geospatial entity types, can be employed in the \texttt{CityFM} framework. In our open-source software, released on GitHub\footnotemark[3], we show how \texttt{CityFM} can be readily used to produce suitable representations of any JSON-formatted spatial entity, from any source. In our experiments, such representations demonstrated to result in superior performance in an array of downstream tasks, spanning urban sensing, geospatial data management and city digital twins, when compared to existing geospatial foundation models, such as SpaBERT \cite{spabert}, general-purpose embeddings, such as GeoVectors \cite{GeoVectors}, and even methods that were tailored for the specific tasks.

\subsection{Predicted Impact}

We envision that \texttt{CityFM}, and future iterations of pre-trained foundation models, will have a broad impact in geospatial applications. 

\subsubsection{Urban Applications}
First, the quality of the entities' embeddings heavily affects the performance of machine learning models employed in downstream urban task. For instance, in next POI recommendation \cite{yile_next_place, geo_information_retrieval, time_aware_poi}, suitable semantic and positional encodings are pivotal to produce relevant results. In addition, \texttt{CityFM} advances the state-of-the-art in road- (polyline) and building- (polygon) based tasks, which can produce a positive impact in many applications, such as population estimation, property price inference and travel time estimation. In Section \ref{sec:traffic_speed}, we show how the road segments' representations can be used to infer traffic speed with better accuracy compared to application-specific algorithms. 

\begin{figure*}
  \includegraphics[width=0.7\textwidth]{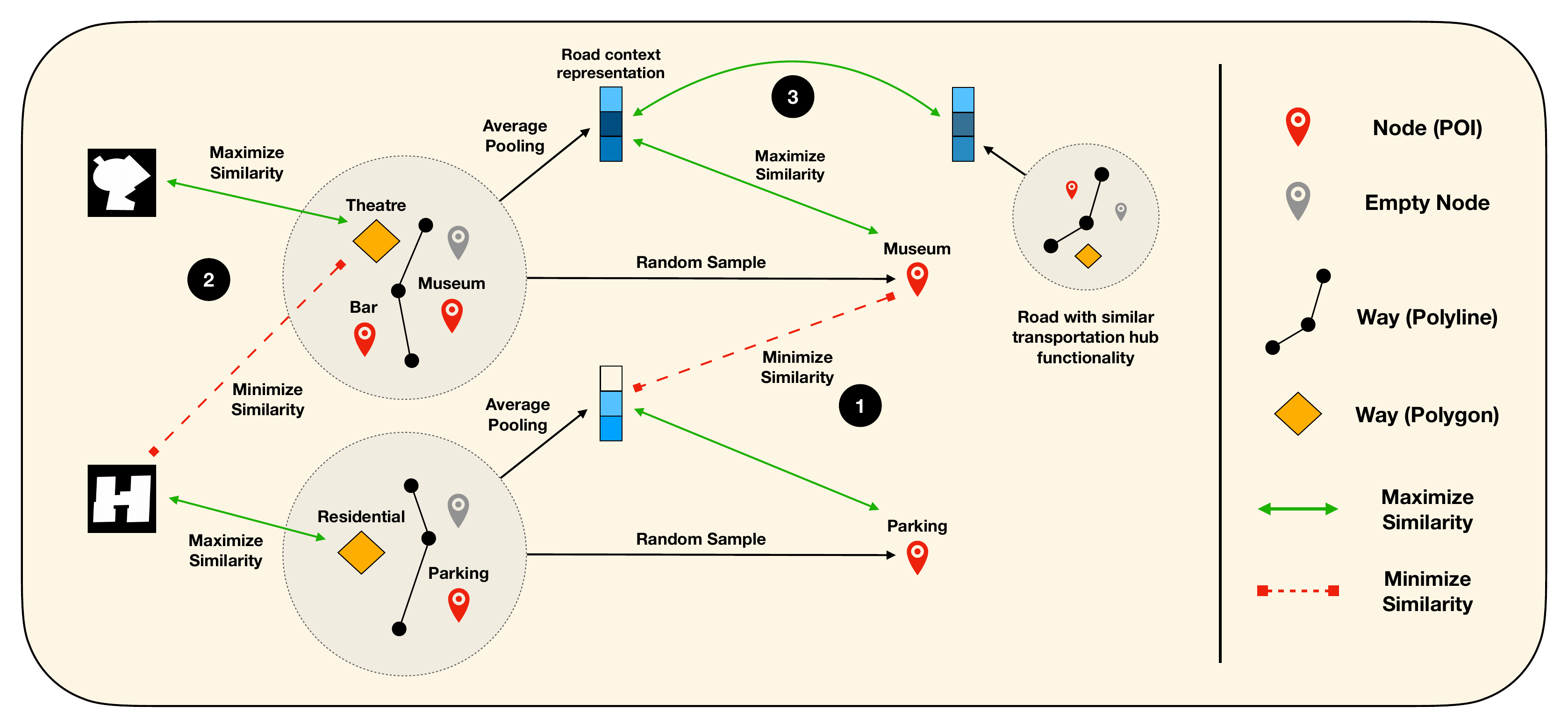}
  \caption{\texttt{CityFM}'s self-supervised pre-training framework. The three contrastive objectives are highlighted: (1) Text-based objective; (2) Vision-Language multimodal objective; (3) Road-based objective. Some dashed red lines are omitted for clarity.}
  \label{framework_figure}
\end{figure*}

\subsubsection{Geospatial Data Management}
\texttt{CityFM} foundation models can ease the process of navigating large geospatial databases, with different entity types, providing universal representations. This can support database applications such as region search \cite{region_search}, geographic information retrieval \cite{geo_entity_querying, top_k_spatial_objects, keyword_search}, and data integration \cite{geo_er}, by reducing the feature extraction cost
. For instance, in data integration, algorithms are typically designed to learn an embedding for each object, to represent similarity between them, and be used to compute a matching probability \cite{geo_er}. The PFMs introduced in this paper, generate meaningful and multimodal representations for various object types and positions, that can be readily used to query similar objects. 
Similarly, Geographic Knowledge Graphs (GeoKGs) \cite{mining_geospatial_rel_from_text} can benefit from the representations produced by our models: the high-dimensional embedding of geographic coordinates, can, in fact, 
lift the quality of the node representation in the KG.

\subsubsection{City Digital Twin}

Enabling Smart Cities through Digital Twins is a promising path \cite{digitial_twin}. Being natively multimodal, our models are capable of fusing the wealth of data from different sensors, to provide us with a holistic understanding of our cities. We design a specific task to demonstrate the utility of our visual representations, to predict the functionality of buildings in the city. 
The functionality of a building is a crucial feature to predict, for instance, its energy consumption \cite{energy_consumption}. The embeddings produced by \texttt{CityFM} for the different types of entities, can be regarded as a digital replica of the spatial objects in the real world, and can be leveraged by experts for urban planning and land use optimization. 

\section{Preliminaries}
\label{sec:preliminaries}
In this section, we provide a detailed description of the data used in this study, and a formal problem definition.
\smallskip

\textit{Definition 2.1} (\texttt{Node}): \textit{A \texttt{Node} $n$ is a point with a geographic position $n.p = (lat, long)$. It can be associated with a set of tags $n.t = \{t_1, ..., t_m\}$, to represent a Point of Interest.} If $n.t = \varnothing$ the node is typically used to construct a more complex geometry, e.g., a polygon, and thus it is not considered as an independent entity.

\smallskip
\textit{Definition 2.2} (\texttt{Way}): \textit{A \texttt{Way} $w$ is associated with an ordered list of nodes $w.n = [n_1, ..., n_k]$, that defines its shape as a polyline. If $n_1 = n_k$, $w$ is a closed polygon. Similar to nodes, a \texttt{Way} can be optionally associated with a set of tags.} Ways of type polyline are used to represent linear features such as roads, bridges and rivers. Ways of type polygon typically represent larger POIs, such as buildings, universities and airports.

\smallskip
\textit{Definition 2.3} (\texttt{Relation}): \textit{A \texttt{Relation} $r$ is associated with an ordered list of entities $r.m = [e_1, ..., e_r]$, called members, which can be a combination of nodes, ways and relations. A \texttt{Relation} can be associated with a set of tags, and each of its members can be associated with a string, defining its role in the relation.} An example of \texttt{Relation} is a bus loop, where a set of polylines (Ways) defines its path, and the bus stops are POIs (Nodes).


\smallskip
\textit{Definition 2.4 (Geospatial Foundational Pre-training): Given a target geographical area (e.g., a city), we aim to pre-train a general model, using OpenStreetMap's geospatial entities in the target area, in a self-supervised fashion. In order to be considered acceptable, such model is expected to generate spatial entity representations that are suitable for use in geospatial downstream tasks.}

\section{CityFM}

\texttt{CityFM} is a framework that pre-trains a geospatial foundation model, using OpenStreetMap's entities in a target region of interest. 
The workflow involves a data preprocessing part, that removes personal information, including phone numbers, URLs and addresses.

\subsection{Self-Supervised Learning}
\label{section:selfsupervised-learning}
In this section, we introduce and provide the underlying motivations for the \texttt{CityFM} framework, depicted in Figure \ref{framework_figure}. As illustrated in Section \ref{sec:preliminaries}, large-scale geospatial databases, such as OpenStreetMap, encompass diverse entity types and data modalities. While the available information is abundant, textual annotations, such as the functionality of a building or the surface type of a road, are sparsely distributed and vary in their representation, due to the lack of an underlying ontology. This significantly reduces the amount of training data available, and the capability of deep learning algorithms to learn meaningful representations.

The purpose of the self-supervised learning phase, is to pre-train a general model to learn a unified representation of the spatial objects, using the different entity types and information modalities available. We design three different contrastive objectives, using nodes, polylines, polygons and relational information. The first objective is a mutual information-based text-to-text objective, and is used to train a language model for representation learning of the textual part of the entities. The second one is a vision-language contrastive objective, whose purpose is to learn a visual representation of an object's shape, that can indicate its functionality. The third one is a road-based context-to-context objective, that leverages public transportation information, found in entities of type \texttt{Relation}, to identify road segments with similar functionalities.

\subsubsection{Text-based Contrastive Objective} A crucial part of a geospatial entity's textual annotation is the information that denotes its functionality, and a common approach in the field is to partition categories into one-hot-encoded classes \cite{hex2vec, opensentinelmap}. In OSM, this is challenging as tags do not follow a pre-defined, structured ontology. For instance, a node could be tagged with the key-value pair "\textit{shop: florist}", which is very specific and the key \textit{shop} alone would provide sufficient information. On the other hand, in a polygon tagged as "\textit{amenity: place\_of\_worship}", the key \textit{amenity} is too generic, and it is used for different classes such as restaurants and hospitals. In addition, certain unique keys or values, devised by individual contributors, are too sparsely used, which hinders the model's ability to learn a meaningful representation. Because of this, we used a pre-trained LM, i.e. BERT \cite{BERT}, to provide an initial representation for the textual part of the entity. Formally, given an entity $e$, associated with a set of tags $e.t = \{t_1, ..., t_m\}$, where each tag $t_i$ is a key-value pair $k_i:v_i$, its high-dimensional representation $\boldsymbol{h}_e$ is computed using BERT and a 2-layer MLP to map it into a multimodal space:


\begin{table}
    \caption{A qualitative comparison between BERT and CityFM (following text-based contrastive pre-training).}
    \label{tab:text_qualitative_results}
    \begin{tabularx}{\linewidth}{lXc}
        \toprule
        &\textbf{BERT}&\\
        \midrule
        $Entity_1$&$Entity_2$&\scalebox{0.85}{Cosine Similarity}\\
        \midrule
        $amenity : hospital$&$amenity : hospital$&1.0\\[0.0cm]
        &$amenity : cafe$&0.91\\[0.0cm]
        &$amenity : restaurant$&0.90\\[0.0cm]
        &$amenity : doctors$&0.87\\[0.0cm]
        & ... &\\[0.0cm]
        &$building : residential$&0.78\\[0.0cm]
        &$healthcare : clinic$&0.77\\[0.0cm]
        \bottomrule
        \toprule
        &\textbf{CityFM}&\\
        \midrule
        $Entity_1$&$Entity_2$&\scalebox{0.85}{Cosine Similarity}\\
        \midrule
        $amenity : hospital$&$amenity : hospital$&1.0\\[0.0cm]
        &$healthcare : clinic$&0.95\\[0.0cm]
        &$amenity : doctors$&0.89\\[0.0cm]
        &$healthcare : pharmacy$&0.79\\[0.0cm]
        &$shop : medical\_supply$&0.74\\[0.0cm]
        & ... &\\[0.0cm]
        &$building : greenhouse$&-0.22\\[0.0cm]
        &$tourism : attraction$&-0.23\\[0.0cm]
        \bottomrule
    \end{tabularx}
\end{table}

\vspace{-3mm}
\begin{center}
\begin{equation}\label{eq:1}
\boldsymbol{h}_e = MLP(Avg(BERT(S_{e.t})))
\end{equation}
\end{center}
$Avg$ is an average-pooling layer to aggregate the BERT representations of the words in the sequence. $S_{e.t}$ is the serialized sequence of \textit{comma-separated} tags, following the standard format for BERT model:

\begin{center}
$S_{e.t}$ = [CLS] $t_1$, $t_2$, ..., $t_m$ [SEP]
\end{center}
\smallskip
The purpose of the first contrastive objective is to push the embeddings of the entities closer in the high-dimensional space, if they are observed in spatial proximity. To achieve this, we group entities located on the same road segment 
and maximise the similarity between a randomly sampled entity and its context, computed as the average of the entities in the same group. This particular choice is motivated by the expectation that entities situated within the same road segment, exhibit stronger correlations \cite{same_road_pois}. Given the fact that the entities in a road context are aggregated with an average-pooling layer, we add an \textit{empty node} in each group, to represent emptiness in the context. Such node has a single tag $n.t = \{context: none\}$. We employ a noise contrastive estimation (NCE) \cite{NCE_loss} loss function, where negative samples are contexts of entities sampled from other groups, in the same minibatch:

\begin{figure*}
  \includegraphics[width=0.6\textwidth]{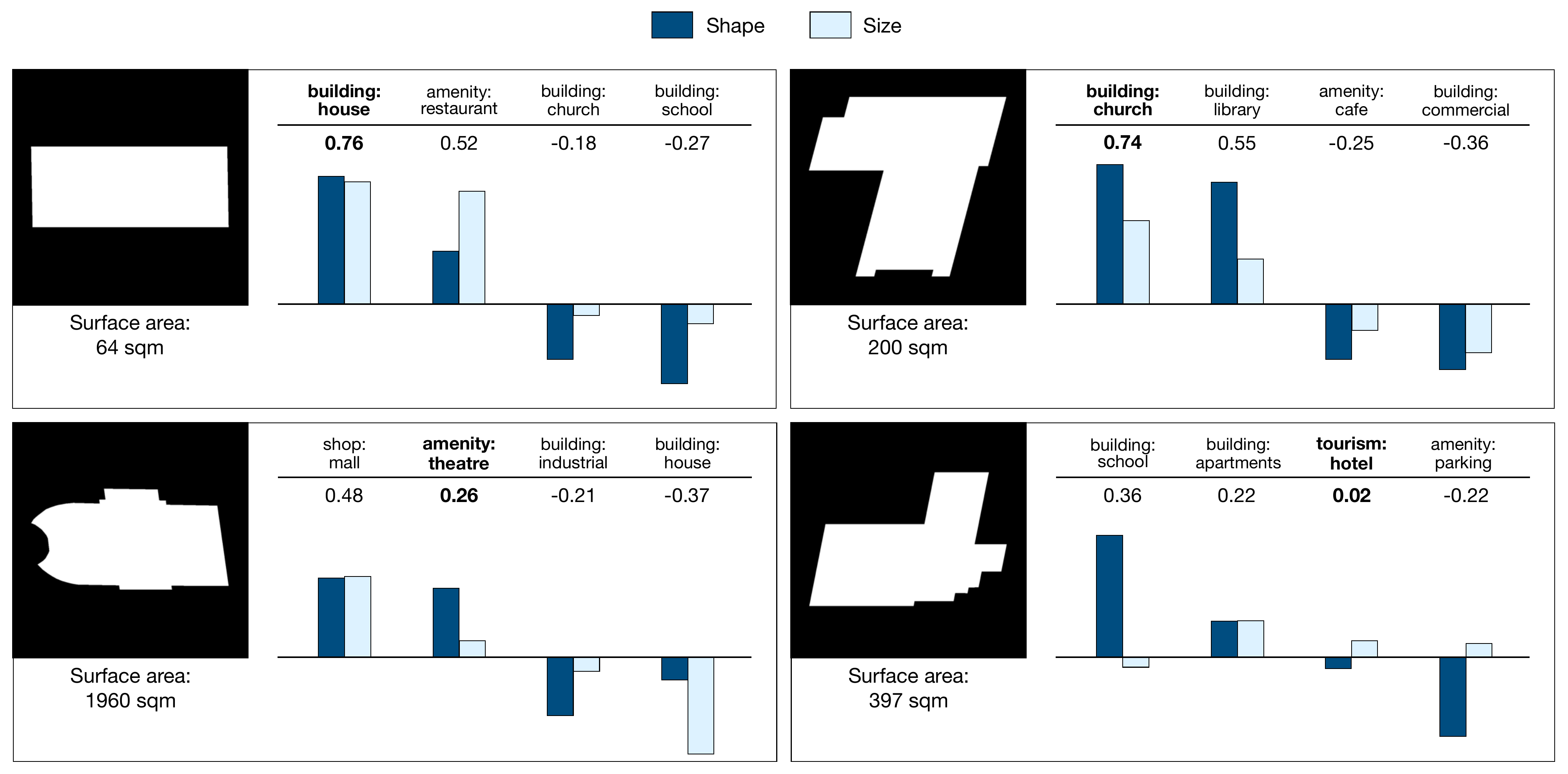}
  \caption{Some examples demonstrating \texttt{CityFM}'s capability to associate the visual characteristics of OpenStreetMap's polygons, with their corresponding functionality. We report the similarities of the \textit{shape} and \textit{size} with the textual encodings of the tags.}
  \label{im:vision_language}
\end{figure*}

\vspace{-3mm}
\begin{center}
\begin{equation}\label{eq:2}
\mathcal{L}_{NCE_T} = - \frac{1}{B} \sum_i^B{log(\frac{exp(\frac{\boldsymbol{h}_i^\top\boldsymbol{c}_i}{\tau})}{\sum_j^B{exp(\frac{\boldsymbol{h}_i^\top\boldsymbol{c}_j}{\tau})}})}.
\end{equation}
\end{center}
\vspace{1mm}
In Eq. \ref{eq:2}, $B$ is the minibatch size,  $\boldsymbol{h}_i$ is the initial representation for entity $e_i$, computed as in Eq. \ref{eq:1}, $\tau$ (= 0.5) is a temperature parameter, and $\boldsymbol{c}_{i}$ is the representation of $\mathcal{C}_{i}$, the context entity $e_i$:
\begin{center}
\begin{equation}\label{eq:3}
\boldsymbol{c}_{i} = Avg(\{\boldsymbol{h_e} : e \in \mathcal{C}_{i}\}) 
\end{equation}
\end{center}
\vspace{1mm}
We train both the language model and the MLP, to minimize $\mathcal{L}_{NCE_T}$.

Table \ref{tab:text_qualitative_results} shows a qualitative comparison between BERT and \texttt{CityFM}. Specifically, we report the cosine similarity between the embedding of some pairs of entities, using their categorical tags. We notice that the similarity of BERT representations is heavily affected by the number of words that the entities share, such as \textit{amenity}. In contrast, \texttt{CityFM}, following the text-based contrastive pre-training, is capable of capturing a deeper semantic similarity, based on the spatial co-occurrence of the entities. As expected, entities that are frequently observed in isolation, such as $\{amenity: fuel\}$, demonstrate significantly higher similarity with $\{context: none\}$.

\subsubsection{Vision-Language Contrastive Objective} While a large amount of entities, in the OpenStreetMap's database, is represented as a polygon, only a small fraction ($\sim$20\%) is associated with tags. Such \textit{untagged} entities have often been considered irrelevant in existing studies \cite{ambiguity_and_plausibility, osm_pois_urban_change, hex2vec}, and therefore discarded. Nonetheless, valuable insights about a building's functionality can be provided by auxiliary information, such as its shape and size. Motivated by this, we design a \textit{cross-modal} contrastive learning objective, during which our model is trained to produce a representation for a polygon's shape, that is as close as possible to the representation of its functionality, in high-dimensional space. During this pre-training phase, we utilize only polygons associated with tags
. Formally, given an entity $e$, of type \texttt{Way}, associated with an ordered list of nodes $e.n = [n_1, ..., n_k]$, where $n_1 = n_k$, and a set of tags $e.t = \{t_1, t_2, ..., t_m\}$, we compute a high-dimensional representation of its shape,

\vspace{-3mm}
\begin{center}
\begin{equation}\label{eq:4}
\boldsymbol{s}_e = MLP(ResNet18(Raster(e.n))),
\end{equation}
\end{center}
where $Raster$ is a \textit{rasterization} function that maps a closed polygon, represented as an ordered list of nodes, to a binary image; $ResNet18$ \cite{resnet} is a pre-trained vision algorithm that computes a high-dimensional representation of an input image. Since the rasterized polygon covers a fixed portion of the image, independently of its size, we separately compute and embed the surface area of the building as follows:

\vspace{-3mm}
\begin{center}
\begin{equation}\label{eq:5}
\boldsymbol{a}_e = MLP(\frac{Surface(e.n)}{max_a}).
\end{equation}
\end{center}
In Eq. \ref{eq:5}, $Surface$ is a function to compute the surface area of a polygon in $m^2$, and $max_a$ is the maximum area of a polygon, used as a normalization constant. The visual representation of the building, $\boldsymbol{v}_e$, is subsequently computed as the arithmetic mean of its shape and size representations, and a 2-layer MLP is used to map it into the multimodal space:

\vspace{-3mm}
\begin{center}
\begin{equation}\label{eq:6}
\boldsymbol{v}_e = MLP(\frac{\boldsymbol{s}_e + \boldsymbol{a}_e}{2}).
\end{equation}
\end{center}
Finally, the following contrastive loss function is used to maximise the similarity, in high-dimensional space, between a building's textual representation $\boldsymbol{h}_e$, computed as in Eq. \ref{eq:1}, and its visual representation $\boldsymbol{v}_e$, while minimizing the similarity with the visual representations of other polygons in the same minibatch:

\begin{center}
\begin{equation}\label{eq:7}
\mathcal{L}_{NCE_V} = - \frac{1}{P} \sum_i^P{log(\frac{exp(\frac{\boldsymbol{h}_i^\top\boldsymbol{v}_i}{\tau})}{\sum_j^P{exp(\frac{\boldsymbol{h}_i^\top\boldsymbol{v}_j}{\tau})}})},
\end{equation}
\end{center}
where $P$ is the number of polygons in the minibatch, and $\tau$ (= 0.5), is a temperature parameter.

In Figure \ref{im:vision_language}, we showcase \texttt{CityFM}'s capability to associate the visual characteristics of OpenStreetMap's polygons\footnote{The polygons in Fig. \ref{im:vision_language} are \textit{tagged} buildings that have been left out during pre-training, for experimental purposes}, with their corresponding functionality. The figure illustrates the similarity of the polygons to some categorical tags in OSM, with the ground truth highlighted in bold font. In addition, we analyze the contribution of the shape and size of the building, by reporting the similarity of their individual representations. 

\subsubsection{Road-based Contrastive Objective}
\begin{figure}
  \centering
  \includegraphics[width=0.8\linewidth]{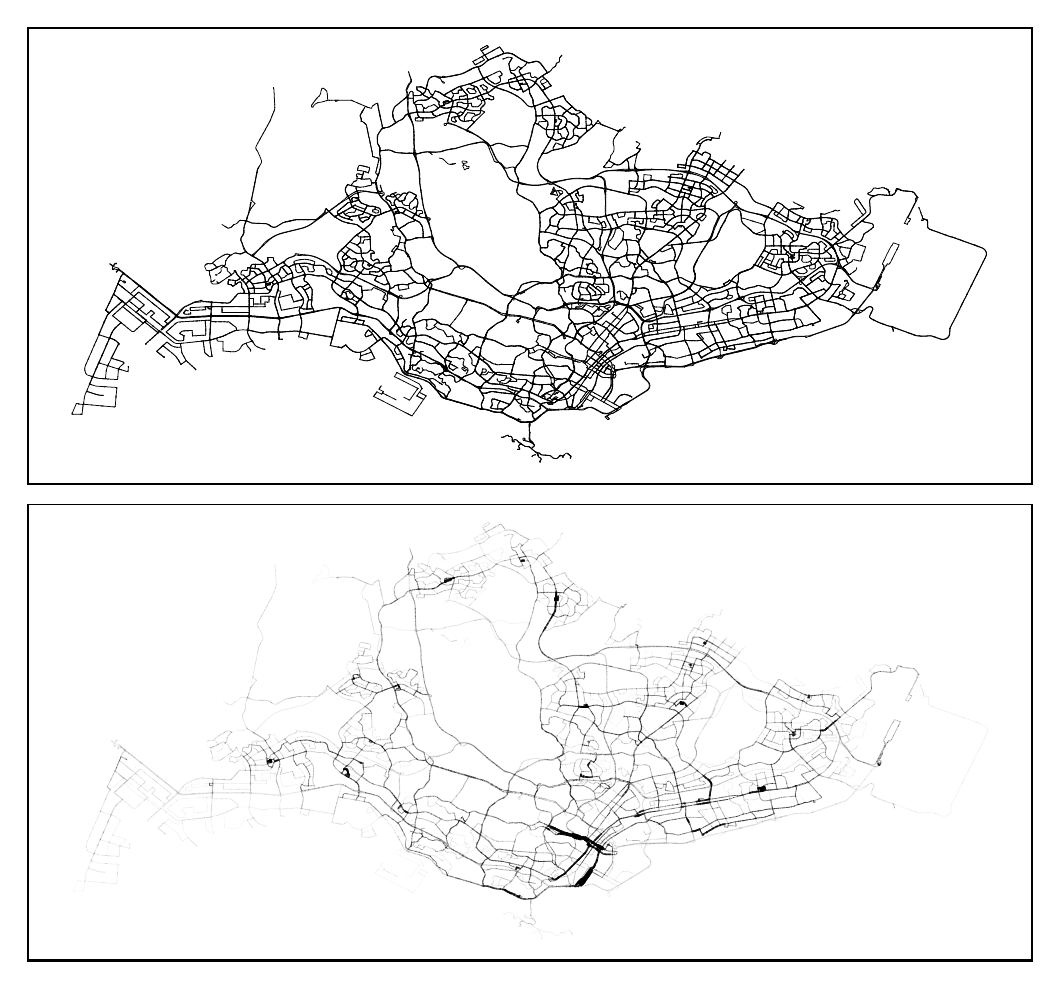}
  \caption{Top: All the road segments traversed by at least one bus loop, in OSM Singapore. Bottom: The same road segments, weighted by number of bus loops traversing them.}
  \label{img:relations}
\end{figure}

As defined in Section \ref{sec:preliminaries}, OSM objects of type \texttt{Relation} are characterized by an ordered list of entities, called members, that together describe a more complex object. Relations mostly represent public transportation routes, such as bus loops and train lines. Although public transportation covers the vast majority of urban areas, some roads represent critical links between transportation hubs and key regions of the city. Such roads, often referred to as arterial roads, experience higher traffic volumes and are traversed by a larger number of public means, compared to others. Figure \ref{img:relations} (top) illustrates all the road segments (polyline ways) in the city of Singapore that are members of at least one relation tagged as "\textit{route: bus}", which implies that at least one bus loop traverses the road. In the bottom image of the figure, each road segment is weighted by the number of bus loops that traverse it. Formally, given an entity $e$, of type \texttt{Way}, associated with an ordered list of nodes $e.n = [n_1, ..., n_k]$, where $n_1 \neq n_k$,
its transportation link weight $l_e$ is computed as the number of relations containing $e$ as a member:

\begin{center}
\begin{equation}\label{eq:8}
l_e = |\{r: e \in r\}|,
\end{equation}
\end{center}

\begin{center}
\begin{equation}\label{eq:9}
l_e = \frac{l_e}{\max_{j\in W}l_j}.
\end{equation}
\end{center}

In Eq. \ref{eq:9}, $l_e$ is normalised, and $W$ is the set of all \texttt{Way} entities. The second picture in Fig. \ref{img:relations} facilitates the identification of the main transportation hubs in Singapore, and highlights arterial roads that connect different areas of the city. Due to the absence of human mobility data in OSM, we decide to use relations to further refine the representation of road segments, initially based on context alone. Specifically, the embedding $\boldsymbol{h}_r$ of a road segment $r$, associated with a transportation link weight $l_r$, is pushed closer, in high-dimensional space, to the embedding of another road with similar transportation link functionality, independently of their distance. In this context, two road segments $r_1$ and $r_2$ are considered similar if the difference between their link weights is smaller than a given threshold $\theta$ (= 0.05). The set of road segments with similar transportation link functionality to $r_e$, is therefore defined as:

\begin{center}
\begin{equation}\label{eq:10}
sim(r_e) = \{ j : |\phantom{|} l_{r_e} - l_{r_j} | < \theta\}
\end{equation}
\end{center}
The following loss function is minimized during the road-based contrastive objective,
\begin{center}
\begin{equation}\label{eq:11}
\mathcal{L}_{NCE_R} = - \frac{1}{R} \sum_i^R{\frac{1}{|sim(r_i)|}\sum_{j \in sim(r_i)}{log(\frac{exp(\frac{\boldsymbol{s}_i^\top\boldsymbol{s}_j}{\tau})}{\sum_{k}^N{exp(\frac{\boldsymbol{s}_i^\top\boldsymbol{s}_k}{\tau})}})}},
\end{equation}
\end{center}
where R is the number of road segments, $\boldsymbol{s_e}$ is the high-dimensional representation for a road segment $e$, initialised as in Eq. \ref{eq:3}, and $N$ is the number of randomly sampled negatives, with $k \notin sim(r_i)$.

\subsection{Location Encoding}
A geospatial entity is defined by its inherent characteristic of occupying a specific location on the planet, which is typically represented as a 2-dimensional point $p = (lat, long)$. Recent studies have proposed various location encoders, designed to project point-based locations into high-dimensional vectors. In \cite{incorporating_semantic_similarity}, the authors employ a fixed-size grid approach, mapping entities' positions to grid cells and learning high-dimensional representations for each cell. In \cite{geohash_cikm, geohash_igarss}, GeoHash is utilized to translate positions into strings, representing specific cells on the surface of the Earth, and whose size depends on the algorithm's precision, specified by the user. GeoVectors \cite{GeoVectors} initializes entities' positions with random vectors, and refines them through random walks with distance-dependent transition probabilities, resulting in closer points having similar representations. In Space2Vec \cite{space2vec_mai}, the authors encode point-based locations with sinusoidal functions and a multi-layer perceptron, using several frequencies to form a global code-book of 2-dimensional positions. Finally, SpaBERT \cite{spabert} incorporates the location encoding in the transformer's architecture.

All the aforementioned approaches involve a training phase to learn the parameters of the location encoder. However, considering the vector nature of positional data, where closer positions exhibit higher similarity in the 2-dimensional vector space, we argue that learned parameters are unnecessary. We follow the positional encoding for words in a sequence, introduced in the original Transformer \cite{transformers} architecture, to define a sinusoidal encoder as follows:

\begin{center}
\begin{equation}\label{eq:12}
f(p_l)^{(i)} := \begin{cases}
      sin(\omega_k\cdot p_l) & \text{if $i = 2k$}\\
      cos(\omega_k\cdot p_l) & \text{if $i = 2k + 1$}
\end{cases}
\forall l = 0, 1
\end{equation}
\end{center}
\begin{center}
\begin{equation}\label{eq:13}
\omega_k = \frac{\lambda}{10000^{\frac{2k}{d}}}
\end{equation}
\end{center}
where $d$ (= 128) is the dimension of the encoding space, and $\lambda$ (= 100) is a rescaling factor that we introduce to facilitate the subsequent learning model's ability to capture the subtle differences that occur between positions expressed in latitude and longitude. In Eq. \ref{eq:12}, $p_0$ is the latitude and $p_1$ is the longitude, and the two vectors are concatenated, leading to a $2d$-dimensional location encoding. Finally, the wide range of frequencies ensures that the position is represented at different granularity levels.

\section{Experiments}
The objective of this section is twofold. First, we aim to demonstrate the effectiveness of foundation models pre-trained using \texttt{CityFM}, when applied to downstream applications. We compare against baselines that are specifically designed for each task, and trained using downstream data directly. Second, we seek to showcase the potential of utilizing data solely from Volunteered Geographic Information sources, such as OpenStreetMap, which is freely available and accessible globally, and how it can provide valuable information to effectively address geospatial challenges of various type.

\subsection{Experimental Settings}
\texttt{CityFM} models are pre-trained using OSM data, on the three contrastive objectives presented in Sec. \ref{section:selfsupervised-learning}. We minimize the pre-training loss $\mathcal{L}_{pt}$, which is the sum of the task-specific losses:
\vspace{-2mm}
\begin{center}
\begin{equation}\label{eq:14}
\mathcal{L}_{pt} = \mathcal{L}_{NCE_T} + \mathcal{L}_{NCE_V} + \mathcal{L}_{NCE_R}
\end{equation}
\end{center}
The model is trained until convergence with a learning rate of $10^{-4}$, a batch size of 256, and a linearly decreasing learning rate scheduler with warm-up. While the framework allows for various text and vision models, we use BERT-base-uncased and ResNet-18. The rasterized images are generated with dimensions of 224x224. Following the pre-training phase, the model's parameters are frozen, and \texttt{CityFM} is utilized to generate meaningful representations for the geospatial entities involved in the different downstream applications.

\begin{table*}
  \caption{Results of Traffic Speed Inference, reporting mean and stddev of 10 independent runs. Speed measure is miles per hour.}
  \label{tab:avg_speed_results}
  \begin{tabular}{ccccccccc}
    \toprule
    &\multicolumn{4}{c}{NYC}&\multicolumn{4}{c}{Seattle}\\
    \cmidrule(lr){2-5}
    \cmidrule(lr){6-9}
    Model & RMSE $\downarrow$ & MAE $\downarrow$ & $R^2 \uparrow$ & MAPE $\downarrow$ & RMSE $\downarrow$ & MAE $\downarrow$ & $R^2 \uparrow$ & MAPE $\downarrow$\\
    \cmidrule(lr){1-1}
    \cmidrule(lr){2-5}
    \cmidrule(lr){6-9}
    \multirow{2}{*}{Node2Vec \cite{node2vec}} & 6.82 & 5.31 & 0.38 & 32.22\% & 7.19 & 6.38 & 0.3715 & 31.24\%\\[-0.1cm]
    & \scalebox{0.80}{($\pm$ 0.12)} & \scalebox{0.80}{($\pm$ 0.05)} & \scalebox{0.80}{($\pm$ 0.04)} & \scalebox{0.80}{($\pm$ 0.6\%)} & \scalebox{0.80}{($\pm$ 0.08)} & \scalebox{0.80}{($\pm$ 0.07)} & \scalebox{0.80}{($\pm$ 0.02)}
    & \scalebox{0.80}{($\pm$ 0.1\%)} \\[0.0cm]

    \multirow{2}{*}{GCWC \cite{gcwc}} & 6.74 & 5.2 & 0.4112 & 32.75\% & 7.14 & 5.71 & 0.4437 & 30.95\%\\[-0.1cm]
    & \scalebox{0.80}{($\pm$ 0.04)} & \scalebox{0.80}{($\pm$ 0.04)} & \scalebox{0.80}{($\pm$ 0.04)} & \scalebox{0.80}{($\pm$ 1.2\%)} & \scalebox{0.80}{($\pm$ 0.06)} & \scalebox{0.80}{($\pm$ 0.09)} & \scalebox{0.80}{($\pm$ 0.01)}
    & \scalebox{0.80}{($\pm$ 0.7\%)} \\[0.0cm]

    \multirow{2}{*}{RFN \cite{RFN_speed_prediction}} & 6.45 & 4.83 & 0.46 & 30.1\% & 7.22 & 5.69 & 0.51 & 30.7\%\\[-0.1cm]
    & \scalebox{0.80}{($\pm$ 0.09)} & \scalebox{0.80}{($\pm$ 0.02)} & \scalebox{0.80}{($\pm$ 0.02)} & \scalebox{0.80}{($\pm$ 0.01\%)} & \scalebox{0.80}{($\pm$ 0.02)} & \scalebox{0.80}{($\pm$ 0.02)} & \scalebox{0.80}{($\pm$ 0.0)}
    & \scalebox{0.80}{($\pm$ 0.1\%)} \\[0.0cm]

    \multirow{2}{*}{GeoVectors \cite{GeoVectors}} & 5.21 & 3.92 & 0.6423 & 24.03\% & 6.43 & 4.99 & 0.601 & 25.23\%\\[-0.1cm]
    & \scalebox{0.80}{($\pm$ 0.07)} & \scalebox{0.80}{($\pm$ 0.06)} & \scalebox{0.80}{($\pm$ 0.06)} & \scalebox{0.80}{($\pm$ 0.3\%)} & \scalebox{0.80}{($\pm$ 0.14)} & \scalebox{0.80}{($\pm$ 0.13)} & \scalebox{0.80}{($\pm$ 0.01)}
    & \scalebox{0.80}{($\pm$ 1.5\%)} \\[0.0cm]

    \multirow{2}{*}{IRN2Vec \cite{IRN2Vec}} & 5.02 & 3.78 & 0.66 & 24.3\% & 6.11 & 4.74 & 0.6298 & 24.2\%\\[-0.1cm]
    & \scalebox{0.80}{($\pm$ 0.04)} & \scalebox{0.80}{($\pm$ 0.02)} & \scalebox{0.80}{($\pm$ 0.01)} & \scalebox{0.80}{($\pm$ 0.0\%)} & \scalebox{0.80}{($\pm$ 0.26)} & \scalebox{0.80}{($\pm$ 0.19)} & \scalebox{0.80}{($\pm$ 0.03)}
    & \scalebox{0.80}{($\pm$ 0.1\%)} \\[0.0cm]

    \multirow{2}{*}{\textbf{CityFM}} & \textbf{4.08} & \textbf{3.2} & \textbf{0.7709} & \textbf{19.27\%} & \textbf{4.9} & \textbf{3.79} & \textbf{0.7499} & \textbf{18.18}\%\\[-0.1cm]
    & \scalebox{0.80}{($\pm$ 0.01)} & \scalebox{0.80}{($\pm$ 0.01)} & \scalebox{0.80}{($\pm$ 0.02)} & \scalebox{0.80}{($\pm$ 0.8\%)} & \scalebox{0.80}{($\pm$ 0.05)} & \scalebox{0.80}{($\pm$ 0.04)} & \scalebox{0.80}{($\pm$ 0.02)}
    & \scalebox{0.80}{($\pm$ 0.2\%)} \\[0.0cm]
    
    \bottomrule
\end{tabular}
\end{table*}

\begin{table*}
  \caption{Ablation study on the Traffic Speed Inference task. We restrict our model's access to only one type of information at a time, to assess their impact. In the last row we exclude the Road-based contrastive objective during pre-training.}
  \label{tab:avg_speed_ablation}
  \begin{tabular}{ccccccccc}
    \toprule
    &\multicolumn{4}{c}{NYC}&\multicolumn{4}{c}{Seattle}\\
    \cmidrule(lr){2-5}
    \cmidrule(lr){6-9}
    Model & RMSE $\downarrow$ & MAE $\downarrow$ & $R^2 \uparrow$ & MAPE $\downarrow$ & RMSE $\downarrow$ & MAE $\downarrow$ & $R^2 \uparrow$ & MAPE $\downarrow$\\
    \cmidrule(lr){1-1}
    \cmidrule(lr){2-5}
    \cmidrule(lr){6-9}

    \multirow{2}{*}{CityFM (PE)} & 6.99 & 5.13 & 0.2811 & 31.86\% & 8.72 & 6.26 & 0.2184 & 30.55\%\\[-0.1cm]
    & \scalebox{0.80}{($\pm$ 0.01)} & \scalebox{0.80}{($\pm$ 0.02)} & \scalebox{0.80}{($\pm$ 0.03)} & \scalebox{0.80}{($\pm$ 0.3\%)} & \scalebox{0.80}{($\pm$ 0.04)} & \scalebox{0.80}{($\pm$ 0.05)} & \scalebox{0.80}{($\pm$ 0.01)}
    & \scalebox{0.80}{($\pm$ 0.5\%)} \\[0.0cm]

    \multirow{2}{*}{CityFM (Tags)} & 6.24 & 5.01 & 0.4537 & \textbf{24.28\%} & 6.34 & 4.81 & 0.4945 & \textbf{23.31\%}\\[-0.1cm]
    & \scalebox{0.80}{($\pm$ 0.19)} & \scalebox{0.80}{($\pm$ 0.11)} & \scalebox{0.80}{($\pm$ 0.04)} & \scalebox{0.80}{($\pm$ 0.5\%)} & \scalebox{0.80}{($\pm$ 0.05)} & \scalebox{0.80}{($\pm$ 0.04)} & \scalebox{0.80}{($\pm$ 0.01)}
    & \scalebox{0.80}{($\pm$ 0.5\%)} \\[0.0cm]

    \multirow{2}{*}{CityFM (Context)} & \textbf{5.85} & \textbf{4.76} & \textbf{0.5821} & 24.39\% & \textbf{5.84} & \textbf{4.71} & \textbf{0.609} & 23.97\%\\[-0.1cm]
    & \scalebox{0.80}{($\pm$ 0.01)} & \scalebox{0.80}{($\pm$ 0.01)} & \scalebox{0.80}{($\pm$ 0.02)} & \scalebox{0.80}{($\pm$ 0.8\%)} & \scalebox{0.80}{($\pm$ 0.01)} & \scalebox{0.80}{($\pm$ 0.01)} & \scalebox{0.80}{($\pm$ 0.0)}
    & \scalebox{0.80}{($\pm$ 0.2\%)} \\[0.0cm]

    \multirow{2}{*}{CityFM (Context w/o Road-based c.o.)} & 6.06 & 4.99 & 0.5755 & 24.8\% & 5.92 & 4.73 & 0.6065 & 24.4\%\\[-0.1cm]
    & \scalebox{0.80}{($\pm$ 0.03)} & \scalebox{0.80}{($\pm$ 0.03)} & \scalebox{0.80}{($\pm$ 0.03)} & \scalebox{0.80}{($\pm$ 0.7\%)} & \scalebox{0.80}{($\pm$ 0.02)} & \scalebox{0.80}{($\pm$ 0.03)} & \scalebox{0.80}{($\pm$ 0.02)}
    & \scalebox{0.80}{($\pm$ 0.5\%)} \\[0.0cm]
    
    \bottomrule
\end{tabular}
\end{table*}

\subsection{Traffic Speed Inference}
\label{sec:traffic_speed}
In this \textit{road-based} task, the average speed on each road segment is utilized as the inference objective. The purpose is to evaluate the quality of \texttt{CityFM}'s representation of road segments by predicting the average speed at which taxis move on a road link. 
The data source is Uber Movement, which provides Uber taxi speed data mapped to OSM road segments. 
Road segments with less than 10 speed measurements are filtered out, obtaining 29,755 data points for NYC, and 6,745 for Seattle. 

\subsubsection{Baselines} We compare our approach with a set of state-of-the-art road network representation methods:

\begin{itemize}[leftmargin=*]
  \item \textbf{Node2Vec} \cite{node2vec} learns representations of nodes in a graph by increasing the similarity between node pairs within $n$-hop neighborhoods, obtained through random walks.
  \item \textbf{GCWC} \cite{gcwc} introduces the Graph Convolutional Weight Completion framework, which exploits the road network graph's topology and the correlation among adjacent edges to estimate and fill in the missing weights in the network.
  \item Relational Fusion Network, \textbf{RFN} \cite{RFN_speed_prediction}, is a modified Graph Convolutional Network (GCN), designed for road network settings. 
  \item \textbf{IRN2Vec} \cite{IRN2Vec} learns vector representations for intersections of road networks, utilizing random walks to generate sequences of adjacent intersections.
  \item \textbf{GeoVectors} \cite{GeoVectors} introduces an open corpus of embeddings for OSM entities, which includes tags and location embeddings. 
  \item \textbf{CityFM} utilizes road segment tags, positional encoding,
  and the learned representation based on the entities in the road's context.
  
\end{itemize}

\subsubsection{Performance Analysis}
We report the root mean square error (RMSE), the mean absolute error (MAE), the coefficient of determination ($R^2$) and the mean absolute percentage error (MAPE), with mean and standard deviation of 10 independent runs, in Table \ref{tab:avg_speed_results}. The speed of taxis is measured in miles per hour (mph), and the average speed in NYC is 19.5 mph ($\sigma = 8.19$), while in Seattle is 24.15 mph ($\sigma = 10.18$). \texttt{CityFM} achieves the best results on both the datasets, with a mean absolute error of 3.2 mph in NYC dataset and 3.79 mph in Seattle's. All the baselines, with the exception of GeoVectors and ours, consider the road network as a (un-) directed graph, and use Graph Convolutional Networks. While this approach effectively highlights road connectivity and attribute propagation, it may under-perform in a real world scenario where measurements are limited to only a small subset of road segments, and additional information need to be incorporated. 

The ablation study in Table \ref{tab:avg_speed_ablation}, shows that the contextual view of road segments emerges as a strong predictor for inferring the average speed at which taxis travel. The entities surrounding a road provide valuable information on its usage and traffic patterns, and serve as a replacement for missing polyline annotations. The positional encoding (PE), while being the weakest predictor, encodes road segments' absolute positions, allowing \texttt{CityFM} to capture spatial relationships between them. 

\subsection{Building Functionality Classification}
\label{sec:building_functionality}
As demonstrated in Figure \ref{im:vision_language}, \texttt{CityFM} is capable of associating building shapes to their respective functionalities. This is accomplished through the Vision-Language contrastive objective, which leverages the OSM buildings that have tags associated with them. Considering that a significant portion of OSM polygons remains untagged, this represents an opportunity to use \texttt{CityFM} to annotate untagged entities, or to assist human annotators in doing so. We chose to evaluate the performance of \texttt{CityFM} using Singapore Governmental data\footnote{https://www.ura.gov.sg/maps/?service=MP}, which provides detailed land use information at the level of individual buildings. 
The dataset consists of 64,384 polygons that belong to one of 8 classes. Detailed statistics about the dataset can be found in Table \ref{tab:building_functionality_stats}. We divided it into training (50\%), validation (25\%) and test (25\%) sets. 

\begin{table}
    \caption{Statistics of the Building Functionality Classification dataset. Number of instances and percentage of the total.}
    \label{tab:building_functionality_stats}
    \begin{tabular}{ccc}
        \toprule
        Functionality&N. of Instances&Percentage\\
        \midrule
        Residential& 43,224 & 67.1\%\\[0.0cm]
        Industrial& 10,431 & 16.2\%\\[0.0cm]
        Commercial & 5,190 & 8.1\%\\[0.0cm]
        Commercial \& Residential& 1,645 & 2.5\%\\[0.0cm]
        Educational & 1,427 & 2.2\%\\[0.0cm]
        Civic \& Community Institution & 1,205 & 1.9\%\\[0.0cm]
        Sports \& Recreation & 751 & 1.2\%\\[0.0cm]
        Transport & 511 & 0.8\%\\
        \bottomrule
    \end{tabular}
\end{table}

\begin{table}
  \caption{Results of building functionality classification. The baselines are grouped by data type (visual, textual, spatial) they can process. Best overall performance is in bold, best performance using a specific data type is underlined.}
  \label{tab:building_functionality_results}
  \begin{tabular}{ccccc}
    \toprule
    \scalebox{0.80}{Type} & \scalebox{0.80}{Model} & \scalebox{0.80}{macro-F1} & \scalebox{0.80}{weighted-F1} & \scalebox{0.80}{Accuracy} \\
    \midrule

    \multirow{2}{*}{v}&\multirow{2}{*}{\scalebox{0.90}{ResNet-18 [Frozen]}} & 11.4\% & 51.19\% & 55.33\%\\[-0.05cm]
    && \scalebox{0.80}{($\pm$ 0.4\%)} & \scalebox{0.80}{($\pm$ 0.8\%)} & \scalebox{0.80}{($\pm$ 0.2\%)} \\[0.cm]
    
    \multirow{2}{*}{v}&\multirow{2}{*}{ResNet-18} & 25.09\% & 64.71\% & \underline{67.14\%}\\[-0.05cm]
    && \scalebox{0.80}{($\pm$ 0.7\%)} & \scalebox{0.80}{($\pm$ 1.1\%)} & \scalebox{0.80}{($\pm$ 1.1\%)} \\[0.cm]

    \multirow{2}{*}{v}&\multirow{2}{*}{CityFM-Visual} & \underline{38.26\%} & \underline{70.16\%} & 65.93\%\\[-0.05cm]
    && \scalebox{0.80}{($\pm$ 0.2\%)} & \scalebox{0.80}{($\pm$ 0.3\%)} & \scalebox{0.80}{($\pm$ 0.2\%)} \\[0.0cm]

    \midrule

    \multirow{2}{*}{t}&\multirow{2}{*}{BERT [Frozen]} & 34.02\% & 35.24\% & 43.41\%\\[-0.05cm]
    && \scalebox{0.80}{($\pm$ 0.2\%)} & \scalebox{0.80}{($\pm$ 0.5\%)} & \scalebox{0.80}{($\pm$ 0.6\%)} \\[0.cm]
    
    \multirow{2}{*}{t}&\multirow{2}{*}{BERT} & 37.56\% & 44.19\% & 51.77\%\\[-0.05cm]
    && \scalebox{0.80}{($\pm$ 0.3\%)} & \scalebox{0.80}{($\pm$ 0.7\%)} & \scalebox{0.80}{($\pm$ 0.7\%)} \\[0.cm]

    \multirow{2}{*}{t}&\multirow{2}{*}{CityFM-Textual} & \underline{49.23\%} & \underline{73.42\%} & \underline{69.16\%}\\[-0.05cm]
    && \scalebox{0.80}{($\pm$ 0.6\%)} & \scalebox{0.80}{($\pm$ 0.2\%)} & \scalebox{0.80}{($\pm$ 0.2\%)} \\[0.0cm]

    \midrule

    \multirow{2}{*}{s + t}&\multirow{2}{*}{GeoVectors} & \underline{47.24\%} & 64.18\% & 69.49\%\\[-0.05cm]
    && \scalebox{0.80}{($\pm$ 1.4\%)} & \scalebox{0.80}{($\pm$ 2.1\%)} & \scalebox{0.80}{($\pm$ 1.5\%)} \\[0.cm]

    \multirow{2}{*}{s + t}&\multirow{2}{*}{SpaBERT} & 45.06\% & \underline{78.47\%} & \underline{75.95\%}\\[-0.05cm]
    && \scalebox{0.80}{($\pm$ 1.1\%)} & \scalebox{0.80}{($\pm$ 2.4\%)} & \scalebox{0.80}{($\pm$ 2.7\%)} \\[0.0cm]

    \midrule
    
    \multirow{2}{*}{v + t}&\multirow{2}{*}{CLIP} & 43.03\% & 80.7\% & 74.98\%\\[-0.05cm]
    && \scalebox{0.80}{($\pm$ 1.1\%)} & \scalebox{0.80}{($\pm$ 1.6\%)} & \scalebox{0.80}{($\pm$ 1.2\%)} \\[0.0cm]

    \midrule
    
    \multirow{2}{*}{s + v + t}&\multirow{2}{*}{CityFM} & \textbf{70.1\%} & \textbf{92.75\%} & \textbf{91.93\%}\\[-0.05cm]
    && \scalebox{0.80}{($\pm$ 1.7\%)} & \scalebox{0.80}{($\pm$ 1.2\%)} & \scalebox{0.80}{($\pm$ 1.3\%)} \\[0.0cm]
    
    \bottomrule
\end{tabular}
\end{table}

\begin{table}
    \caption{Category-specific performance.}
    \label{tab:building_functionality_ablation}
    \begin{tabular}{cccc}
        \toprule
        Functionality&GeoVectors&SpaBERT&CityFM\\
        \midrule
        Residential& 76.97\% & \underline{82.15\%} & \textbf{96.03\%}\\[0.05cm]
        Industrial& 70.57\% & \underline{74.25\%} & \textbf{94.65\%}\\[0.05cm]
        Commercial & \textbf{88.22\%} & 66.66\% & \underline{87.07\%}\\[0.05cm]
        Comm. \& Res. & \underline{21.08\%} & 18.14\% & \textbf{46.29\%}\\[0.05cm]
        Educational & \underline{57.98\%} & 46.05\% & \textbf{73.00\%}\\[0.05cm]
        Civic \& Commun. Inst. & \underline{13.82\%} & 12.46\% & \textbf{24.92\%}\\[0.05cm]
        Sports \& Recr. & 31.73\% & \underline{38.62\%} & \textbf{69.76\%}\\[0.05cm]
        Transport & \underline{21.84\%} & 17.09\% & \textbf{61.53\%}\\[0.05cm]
        \bottomrule
    \end{tabular}
\end{table}

\subsubsection{Baselines}
We compare \texttt{CityFM} with baselines that are capable of handling textual, visual and/or geospatial information:

\begin{itemize}[leftmargin=*]
  \item \textbf{BERT} \cite{BERT} is a pre-trained LM that serves as the foundation of \texttt{CityFM}'s textual encoding model. We evaluate its performance in two settings: with frozen weights, where only a 2-layers MLP is trained on top of it, and when fine-tuned on the task.

  \item \textbf{ResNet-18} \cite{resnet} is a pre-trained CNN-based, vision algorithm that serves as the foundation of \texttt{CityFM}'s visual encoding model. 

  \item \textbf{CLIP} \cite{CLIP} is a contrastive Language-Image pre-trained algorithm. We use it to encode polygons' visual characteristics and the textual information from nearby entities. The model is fine-tuned.

  \item We use the \textbf{GeoVectors} \cite{GeoVectors} corpus to retrieve the location encoding of the untagged polygon and the GeoVectors-tags embeddings to represent the entities in its spatial proximity.

  \item We use \textbf{SpaBERT} \cite{spabert} to encode the name attribute of the entities that form the context of the untagged building to be classified. 

  \item \textbf{CityFM} is capable of encoding the visual characteristics of a polygon, its position, and the textual features of nearby entities.
  
\end{itemize}

\subsubsection{Performance Analysis} Table \ref{tab:building_functionality_results} shows the overall best results in bold font, and the best results within a given baseline type underscored. \texttt{CityFM}'s textual and visual components, although not fine-tuned on this task, perform better than algorithms that we trained specifically for the it. ResNet-18, when fine-tuned, achieves higher accuracy. GeoVectors \cite{GeoVectors}, SpaBERT \cite{spabert} and CLIP \cite{CLIP}, being able to process multi-modal information, represent strong baselines. \texttt{CityFM}, is capable of producing meaningful representations for the textual and visual characteristics of geospatial entities, and can encode positions in the space, achieving the best results.

In Table \ref{tab:building_functionality_ablation}, we illustrate the F1-scores of GeoVectors, SpaBERT and CityFM for each category of the building functionality classification task. We specifically chose to compare to GeoVectors and SpaBERT, as they represent direct competitors in utilizing data from volunteered geospatial information sources, and a pre-training approach to generate general purpose representations. The results reveal a performance gap between the two approaches and \texttt{CityFM}. 

\subsection{Population Density Estimation}
\label{sec:population_density}
In this task, our objective is to estimate the average population density in different regions of NYC and Singapore.
The primary goal of this task is to demonstrate that the representations produced by \texttt{CityFM} can be used in aggregation to effectively represent regions. 

\subsubsection{Baselines}
The baselines are region embedding methods:

\begin{itemize}[leftmargin=*]
  \item \textbf{Place2Vec} \cite{place2vec} learns entities' representations based on their spatial co-occurrence, inspired by word2vec's algorithm \cite{word2vec}. 

  \item \textbf{Urban2Vec} was proposed in \cite{urban2vec}, and learns a low-dimensional representation for each urban region utilizing images and POIs from the region itself. 

  \item We retrieve from the \textbf{GeoVectors} \cite{GeoVectors} corpus the locations and tags encodings for all the spatial entities inside a region, and utilize an average aggregation function.

  \item \textbf{HGI} \cite{hgi} uses Hierarchical Graph Infomax to learn representations at the POI- and region-levels. 

  \item We use \textbf{SpaBERT} \cite{spabert} pre-trained LM to encode the entities in each region, and we aggregate them by computing the mean of their representations. 

  \item We utilize the representations that \textbf{CityFM} produces for the textual annotations of the entities, the visual shapes of the polygons located inside a region, and the position of the region itself.
  
\end{itemize}

\begin{table}
  \caption{Estimation of population density.}
  \label{tab:pop_density_results}
  \begin{tabular}{ccccccc}
    \toprule
    &\multicolumn{3}{c}{\scalebox{0.90}{Singapore}}&\multicolumn{3}{c}{\scalebox{0.90}{NYC}}\\
    \cmidrule(lr){2-4}
    \cmidrule(lr){5-7}
    Model & \scalebox{0.80}{RMSE $\downarrow$} & \scalebox{0.80}{MAE $\downarrow$} & \scalebox{0.80}{$R^2 \uparrow$} & \scalebox{0.80}{RMSE $\downarrow$} & \scalebox{0.80}{MAE $\downarrow$} & \scalebox{0.80}{$R^2 \uparrow$}\\
    \cmidrule(lr){1-1}
    \cmidrule(lr){2-4}
    \cmidrule(lr){5-7}
    \multirow{2}{*}{\scalebox{0.90}{Place2Vec}} & 10.09 & 7.1 & 0.17 & 10.95 & 8.49 & 0.25\\[-0.1cm]
    & \scalebox{0.70}{($\pm$ 0.3)} & \scalebox{0.70}{($\pm$ 0.21)} & \scalebox{0.70}{($\pm$ 0.05)} & \scalebox{0.70}{($\pm$ 0.35)} & \scalebox{0.70}{($\pm$ 0.24)} & \scalebox{0.70}{($\pm$ 0.02)} \\[0.0cm]

    \multirow{2}{*}{\scalebox{0.90}{Urban2Vec}} & 5.76 & 4.48 & \underline{0.59} & 6.31 & 5.94 & 0.66\\[-0.1cm]
    & \scalebox{0.70}{($\pm$ 0.04)} & \scalebox{0.70}{($\pm$ 0.04)} & \scalebox{0.70}{($\pm$ 0.04)} & \scalebox{0.70}{($\pm$ 0.06)} & \scalebox{0.70}{($\pm$ 0.09)} & \scalebox{0.70}{($\pm$ 0.01)}\\[0.0cm]

    \multirow{2}{*}{\scalebox{0.90}{GeoVectors}} & 6.38 & 5.2 & 0.51 & 7.56 & 6.09 & 0.58\\[-0.1cm]
    & \scalebox{0.70}{($\pm$ 0.23)} & \scalebox{0.70}{($\pm$ 0.27)} & \scalebox{0.70}{($\pm$ 0.06)} & \scalebox{0.70}{($\pm$ 0.3)} & \scalebox{0.70}{($\pm$ 0.22)} & \scalebox{0.70}{($\pm$ 0.01)} \\[0.0cm]

    \multirow{2}{*}{\scalebox{0.90}{HGI}} & \underline{5.31} & \underline{4.06} & 0.57 & \underline{6.15} & \textbf{4.18} & \textbf{0.72}\\[-0.1cm]
    & \scalebox{0.70}{($\pm$ 0.57)} & \scalebox{0.70}{($\pm$ 0.31)} & \scalebox{0.70}{($\pm$ 0.05)} & \scalebox{0.70}{($\pm$ 0.23)} & \scalebox{0.70}{($\pm$ 0.17)} & \scalebox{0.70}{($\pm$ 0.02)} \\[0.0cm]

    \multirow{2}{*}{\scalebox{0.90}{SpaBERT}} & 6.89 & 4.85 & 0.49 & 8.08 & 5.69 & 0.51\\[-0.1cm]
    & \scalebox{0.70}{($\pm$ 0.68)} & \scalebox{0.70}{($\pm$ 0.43)} & \scalebox{0.70}{($\pm$ 0.05)} & \scalebox{0.70}{($\pm$ 0.19)} & \scalebox{0.70}{($\pm$ 0.16)} & \scalebox{0.70}{($\pm$ 0.03)} \\[0.0cm]

    \multirow{2}{*}{\scalebox{0.90}{\textbf{CityFM}}} & \textbf{4.31} & \textbf{2.75} & \textbf{0.82} & \textbf{5.95} & \underline{4.26} & \textbf{0.72}\\[-0.1cm]
    & \scalebox{0.70}{($\pm$ 0.22)} & \scalebox{0.70}{($\pm$ 0.18)} & \scalebox{0.70}{($\pm$ 0.04)} & \scalebox{0.70}{($\pm$ 0.27)} & \scalebox{0.70}{($\pm$ 0.2)} & \scalebox{0.70}{($\pm$ 0.02)} \\[0.0cm]
    
    \bottomrule
\end{tabular}
\end{table}

\subsubsection{Performance Analysis}
The results are reported in Table \ref{tab:pop_density_results}. Population density is expressed in thousands of people per square kilometer. 
GeoVectors explicitly encodes textual and spatial information of the entities, producing more meaningful representations. 
Urban2Vec only implicitly represents spatial proximity, through its contrastive objective. However, it utilizes both textual and visual information, achieving higher performance. SpaBERT under-performs on this particular task, probably due to its inability to capture the proximity among regions. 
HGI, being a recent algorithm specifically tailored for region-level tasks, performs comparably to \texttt{CityFM} on the NYC's dataset. 

\section{Related Work}
\subsection{Pre-trained Foundation Models}
Despite their success in many domains, there exists very little work exploring the development of PFMs for geospatial artificial intelligence \cite{geoai_survey_weiming}. GeoVectors \cite{GeoVectors} is an open corpus of OSM entities' representations. It is a framework to efficiently produce embeddings for a representative set of snapshots of OSM, using FastText \cite{fasttext} for the textual annotations and random walks to learn a representation of the entities' locations. Given that the embeddings are task-agnostic, this can be regarded as an initial effort towards general-purpose representations of geospatial entities. More recently, the authors of SpaBERT \cite{spabert} proposed a framework to train an LLM, initialized with BERT's weights, using geospatial data. 
SpaBERT is capable of producing representations only for the textual part of OSM nodes. In contrast with \texttt{CityFM}, SpaBERT uses entities' names and does not incorporate their tags. Finally, existing works are not designed to handle different entity types.

\subsection{Downstream Tasks}
\subsubsection{Traffic Speed Inference}
In this task, the objective is to infer the average speed at which the traffic moves on \textit{unseen} roads. 
This problem has been studied extensively by researchers. GCWC \cite{gcwc} utilizes only the topology of the road network through GCNs, and stochastic weight completion to predict edge weights. Relational Fusion Networks (RFNs) \cite{RFN_speed_prediction} were proposed as an alternative to classical GCNs, to specifically model road networks. 
In previous works the road network has been treated as a graph.
In real world scenarios, where measurement are often available for a limited amount of road segments, and road-level features are sparsely distributed, algorithms relying solely on information propagation may not yield satisfactory performance. \texttt{CityFM} considers contextual information by utilizing entities located in proximity of the road, and taking into account the road's absolute spatial position, which helps the model learn how roads are used in the city.

\subsubsection{Building Functionality Classification} In order to evaluate the quality of our embeddings, we chose to classify the functionality of OSM untagged buildings, given that visual, textual and positional information can be leveraged in this task. This represents a step forward towards automatically annotating geospatial databases
. However, this problem has not been studied extensively enough. In \cite{building_functionality_germany}, the authors use machine learning techniques and a set of features to classify buildings of various types. The approach presents several drawbacks, such as the use of OSM tags and building use codes from external sources, as input features. In \cite{lidar_building_type}, researchers used LiDAR features with ML models to to discern between residential and non-residential buildings. Other studies, have harnessed airborne laser scanning data \cite{airborne_laser_scanning} and heat consumption \cite{building_type_heat}. Existing approaches have utilized information from different sources, which are available only in selected areas, and functionality information of different types, such as OSM tags and land use labels.

\subsubsection{Population Density Estimation}
Population density estimation in urban regions has crucial consequences in urban planning and resource allocation, and is an important signal of socioeconomic development. 
Place2Vec \cite{place2vec} is designed to learn POIs categories embeddings, by maximizing the similarity of categories that are often observed in spatial proximity. In contrast, Urban2Vec \cite{urban2vec} utilizes multimodal data: the textual tokens of POIs' tags and street view images are mapped to the same embedding space, using a triplet loss function. Recently, HGI \cite{hgi} was proposed to learn entities' and regions' representations jointly and in a self-supervised fashion. 
Compared to \texttt{CityFM}, existing approaches suffer from two main drawbacks: first, they are tailored specifically for region representation learning and cannot be used in different applications; secondly, they are designed to handle limited multimodal data.

\section{Conclusions}
This study, presents a novel framework to pre-train foundation models, in a target geographical area of interest. This is a step forward towards a wider adoption of general purpose AI algorithms to address urban challenges. \texttt{CityFM} relies exclusively on OSM data, facilitating its adoption and reproducibility. Its multimodal capabilities enable it to access diverse entity types, and learn better representations, using different aspects of each entity, which result in its superior performance. 

\begin{acks}
This work was partially funded by the Knut and Alice Wallenberg Foundation (KAW 2019.0550).
\end{acks}


\bibliographystyle{ACM-Reference-Format}
\bibliography{sample-base}

\end{document}